# Pressure-induced color change in the lutetium dihydride LuH$_2$

Pengfei Shan[1,2#], Ningning Wang[1,2#*], Xiquan Zheng[3], Qingzheng Qiu[3], Yingying Peng[3], and Jinguang Cheng[1,2*]

[1]Beijing National Laboratory for Condensed Matter Physics and Institute of Physics, Chinese Academy of Sciences, Beijing 100190, China

[2]School of Physical Sciences, University of Chinese Academy of Sciences, Beijing 100190, China

[3]International Center for Quantum Materials, School of Physics, Peking University, Beijing 100871, China

# These authors contributed equally to this work.

*Corresponding authors: nnwang@iphy.ac.cn; jgcheng@iphy.ac.cn

**Abstract:** The lutetium dihydride LuH$_2$ is stable at ambient conditions. Here we show that its color undergoes sequential changes from dark blue at ambient pressure to pink at ~2.2 GPa and then to bright red at ~4 GPa upon compression in a diamond anvil cell. Such a pressure-induced color change in LuH$_2$ is reversible and it is very similar to that recently reported in the N-doped lutetium hydride. However, our preliminary resistance measurements on LuH$_2$ under pressures up to 7.7 GPa evidenced no superconductivity down to 1.5 K.

**Introduction**

The recent discovery of high-temperature superconductivity in LaH$_{10}$ [1, 2] with the clathrate structure has reignited great enthusiasm on the explorations of rare-earth hydrides with an aim to achieve room-temperature superconductivity. Among the 14 rare-earth elements, lutetium (Lu) is the heaviest one and has the smallest metallic radius, and it is perhaps the least naturally abundant. Theoretical studies on the Lu−H system have predicted high-temperature superconductivity at relatively low pressures, *e.g.* the *Im-3m* LuH$_6$ with $T_c$ = 273 K at 100 GPa [3]. Experimentally, several Lu hydrides have been reported so far. The Lu dihydride LuH$_2$ with the fluorite structure is stable at ambient conditions [4], and it shows a metallic conductivity [5, 6]. By heating a Lu foil with ammonia borane (NH$_3$BH$_3$) in a diamond anvil cell (DAC) at about 110 GPa, Shao *et al.* have obtained a superconducting LuH$_3$ with the critical temperature in the range of $T_c$ = 12-15 K over the pressure range 110-170 GPa [7]. When the synthesis pressure of the above reaction in DAC is elevated to 184 GPa, Li *et al.* reported the formation of a new polyhydride with the possible formula of Lu$_4$H$_{23}$, which adopts a higher H content and shows higher $T_c$ reaching 65-71 K over the pressure range 180-218 GPa [8]. Surprisingly, Dias and coworkers recently reported that N-doped Lu hydride exhibits room-temperature superconductivity at a remarkably low pressure of 1 GPa [9]. According to this report, the sample was recovered from the reaction of a thin Lu foil with H$_2$/N$_2$ (99:1) mixture in DAC at 2 GPa and 65 °C for overnight. It shows a shinning blue color at ambient pressure. Intriguingly, the sample's color was observed to change from blue to pink in the pressure range 0.3-3 GPa (phase II) and then to red at pressures above 3 GPa (phase III). Moreover, the pink-colored phase II was found to exhibit room-temperature superconductivity with the maximum $T_c$ = 294 K achieved at about 1 GPa. Based on the XRD results, the authors have attributed the superconducting phase II to the major phase of $LuH_{3-\delta}N_\varepsilon$ (92.25%) coexisting with the minor phases of $LuN_{1-\delta}H_\varepsilon$ (7.29%) and $Lu_2O_3$ (0.46%) [9]. These discoveries have immediately aroused worldwide interest and will no doubt stimulate extensive studies on the Lu hydrides. Here, we focus on the ambient-stable dihydride $LuH_2$ and report on the observation of pressure-induced color change upon compression in a fashion similar to that reported by Dias and coworkers [9]. Unfortunately, we did not observe superconductivity in $LuH_2$ under pressures up to 7.7 GPa.

**Results**

We purchased the "Lu" powder (L110933-1g, 99.9% metal basis) from Aladdin. As shown in the inset of Fig. 1, the as-received powder is dark blue in color, whereas the Lu in the pure form is expected to be silvery white. Such a difference motived us to verify its phase purity by measuring the powder XRD at room temperature. It turns out that the as-received "Lu" powder actually is a mixture of $LuH_2$ and $Lu_2O_3$ with only a small portion of Lu. Two additional weak peaks at 37.51° and 54.06° marked by asterisks in Fig. 1 come from some unknown phases, which thus cannot be considered in the Rietveld refinement. Based on the scaling factors of the Rietveld refinement on the XRD patten in Fig. 1, their percentages are estimated to be 69.6(3)%, 27.1(6)%, and 3.3(1)% for $LuH_2$, $Lu_2O_3$, and Lu, respectively. The lattice parameter of $LuH_2$ is calculated to be $a$ = 5.02949(8) Å, which agrees well with the reported value of 5.033 Å in literature [4]. Because the $Lu_2O_3$ powder is white in color, the dark blue color of the as-received "Lu" powder should be attributed to the main phase $LuH_2$.

About 100 mg of the above "$LuH_2+Lu_2O_3+Lu$" mixture powder and 200 mg of $CaH_2$ contained in separated $Al_2O_3$ crucibles are sealed together in a quartz tube and then heated at 200 °C for 5 hours. We found that such a treatment did not change the relative ratio of the components in the mixture. However, a closer inspection of the powder under microscope reveals the presence of some shinning grains in the blue color and with dimensions about 40-50 μm, as seen in the inset of Fig. 2(a). Interestingly, the color of these grains is very similar to that of N-doped Lu hydride reported by Dias and coworkers [9]. We then selected several grains and conducted XRD measurements by using a custom-designed instrument equipped with a Xenocs Genix3D Mo Kα (17.48 keV) X-ray source. This source produces a beam spot size of 150 μm at the sample position and provides $2.5 \times 10^7$ photons/sec. These grains shown in the inset of Fig. 2(a) were attached to a copper (Cu) base and then mounted on a Huber 4-circle diffractometer. Diffraction signals are collected by a highly sensitive single-photon counting PILATUS3 R 1M solid state area detector with 981 × 1043 pixels. Each pixel size is 172 $\mu$m × 172 $\mu$m. The XRD pattern was measured in the reflection geometry, with the incident angle rotating from 1° to 31°.

The top XRD pattern in Fig. 2(a) contains the signals from both the samples and the Cu base, while the bottom profile displays only the diffraction from the Cu base. The observed double peaks arise from the Mo $K\alpha_1$ ($\lambda$ = 0.7903 Å) and $K\alpha_2$ ($\lambda$ = 0.7136 Å) lines. Excluding the diffraction peaks from the Cu base, the remaining peaks match well with the calculated ones of LuH$_2$ with the cubic *Fm-3m* space group and lattice parameter $a$ = 5.02949 Å. We noted that the obtained lattice parameter is very close to that of $a$ = 5.0289(4) Å for the Compound A mentioned in Ref. [9]. In the LuH$_2$ structure, the Lu atom is bonded in a body-centered cubic geometry to eight equivalent H atoms with the Lu-H bond lengths being 2.1778 Å.

For comparison, we prepared the Lu trihydride LuH$_3$ by treating the Lu and NH$_3$BH$_3$ at 4 GPa and 1000 °C for 30 min in a large-volume Kawai-type multianvil module. From the XRD measurement at ambient conditions, the calculated lattice parameter $a$ = 5.16295(3) Å of LuH$_3$ is much larger than that of LuH$_2$.

We also collected the Raman spectra of the LuH$_2$ grains at ambient condition by using a confocal Raman system (MonoVista CRS+ 500, Spectroscopy & Imaging GmbH) with 532 nm laser excitation. The obtained spectra are displayed in Fig. 2(b) and compared with those of N-doped Lu hydride given in Ref. [9] (Extended data Fig. 1). For LuH$_2$, we observed pronounced Raman peaks at 250 and 1200 cm$^{-1}$, which are identical to those of N-doped Lu hydride [9]. However, some obvious differences are observed in the spectra below 160 cm$^{-1}$.

After obtaining and isolating these blue grains of LuH$_2$, we loaded one piece of grain into the DAC of 300 μm culet and monitored its color change upon compression at room temperature. The pressure inside DAC was determined from the shift of Ruby fluorescence lines at room temperature, and the soft KBr was employed as the pressure transmitting medium. Figure 3 shows the obtained microphotographs of LuH$_2$ at different pressures. As can be seen, the blue color of LuH$_2$ first changes to pink at 2.23 GPa, and then to bright red at pressures above ~ 4 GPa. After releasing pressure from 5.7 GPa to ambient, its color returns to the original dark blue, confirming that the phase transformation upon compression is reversible. These observations are stunningly similar to those of the N-doped Lu hydride reported by Dias and coworkers [9], even though the critical pressures for the color changes are slightly different. Such a difference might arise from the distinct sample forms in these two studies, *e.g.* the fine powder in Ref. [9] versus the dense grain in the present study. Nonetheless, our present results demonstrate unambiguously that LuH$_2$ undergoes pressure-induced color change in a moderate pressure range. The origin of such phenomenon deserves further investigations.

LuH$_2$ shows the metallic behavior at ambient pressure. To check how the transport properties of LuH$_2$ evolve with pressure accompanying the color changes, we further measured its temperature-dependent resistance under different pressures by employing the four-probe method in DAC with 300 μm culet. A piece of LuH$_2$ grain was loaded into the gasket hole filled with soft KBr as pressure transmitting medium, and four Pt electrodes were put into contact directly with the sample. The pressure was calibrated at room temperature by using the ruby fluorescent method as done above. The temperature-dependent experiments from 300 to 1.5 K were carried out in a liquid-helium cryostat equipped with a resistance heater on the sample mount for the warming-up control.

Figure 4 shows the obtained resistance $R(T)$ curves at three pressures. The inset photographs further confirm the color changes upon compression. For each measurement, we have given in Fig. 4 the pressure values at room temperature before cooling down and after warming up. As can be seen, the pressure always increases after one thermal cycle. In addition, the pressure value may also change significantly upon varying temperatures due to the thermal contraction of the DAC body. These factors render some strange up-down features, relatively low data quality, and irreversibility of cooling-down/warming-up $R(T)$ curves, especially at lower pressures and/or at the high-temperature region. For these reasons, we will not consider the detailed features of these $R(T)$ curves but extract only the immediate messages from our results: (1) $LuH_2$ remains in the metallic state with a relatively low resistance, and (2) superconductivity was not observed down to 1.5 K in the studied pressure range. In addition, it is noted that the absolute resistance first decreases from 0.7 to 3 GPa and then increases from 3.7 to 6.7 GPa. A weak upturn in $R(T)$ appears in the low temperature region at 6.7 GPa. Measurements of $R(T)$ under better hydrostatic pressure conditions are needed in the future in order to obtain reliable information about the pressure effect on the transport properties of $LuH_2$.

**Conclusion**

In summary, we have successfully isolated large grains of Lu dihydride $LuH_2$ and demonstrated that its color changes consecutively from blue at ambient pressure to pink at 2.23 GPa and then to bright red at ~ 4 GPa. Such a pressure-induced color change in $LuH_2$ is reversible and it is very similar to that observed by Dias and coworkers in the N-doped Lu hydride. Unfortunately, our preliminary resistance measurements on the $LuH_2$ sample under pressures up to 7.7 GPa reveal the absence of superconductivity down to 1.5 K.

**Acknowledgements**

We are grateful to Prof. Zhongxian Zhao for his encouragement and continuous support. This work is supported by the National Natural Science Foundation of China (Grant Nos. 12025408, 11921004, 11888101, 11974029), the Beijing Natural Science Foundation (Z190008), the National Key R&D Program of China (2018YFA0305700, 2021YFA1400200), the Strategic Priority Research Program of CAS (XDB33000000), and the CAS Interdisciplinary Innovation Team (JCTD-2019-01).

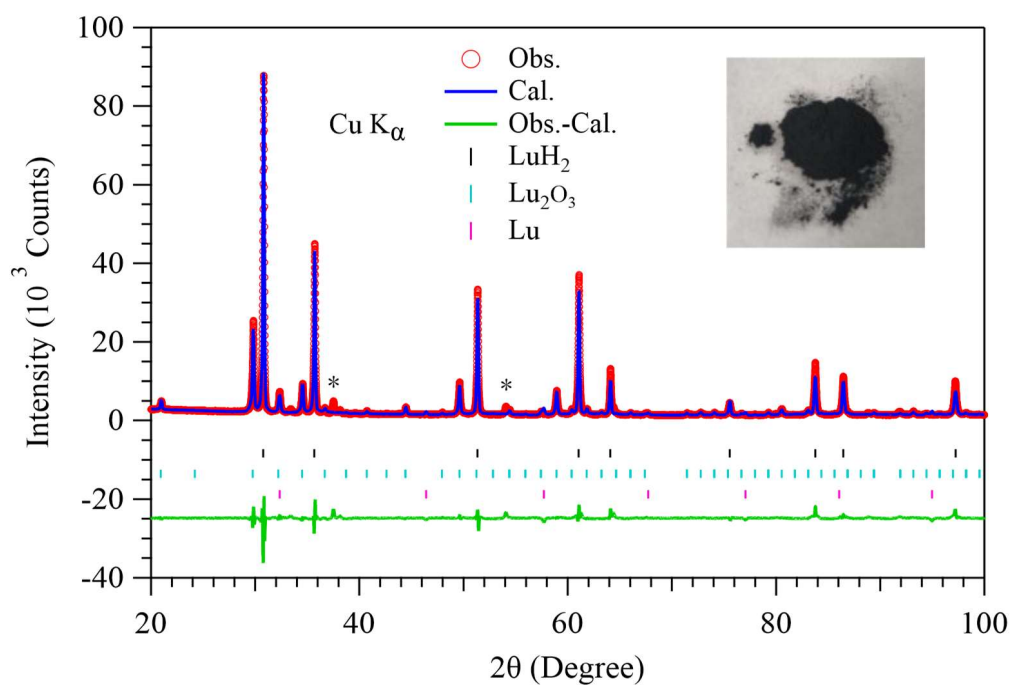

**Figure 1.** Rietveld refinement on the XRD pattern of as-received "Lu" powder from Aladdin. Inset shown the photograph of the "Lu" powder.

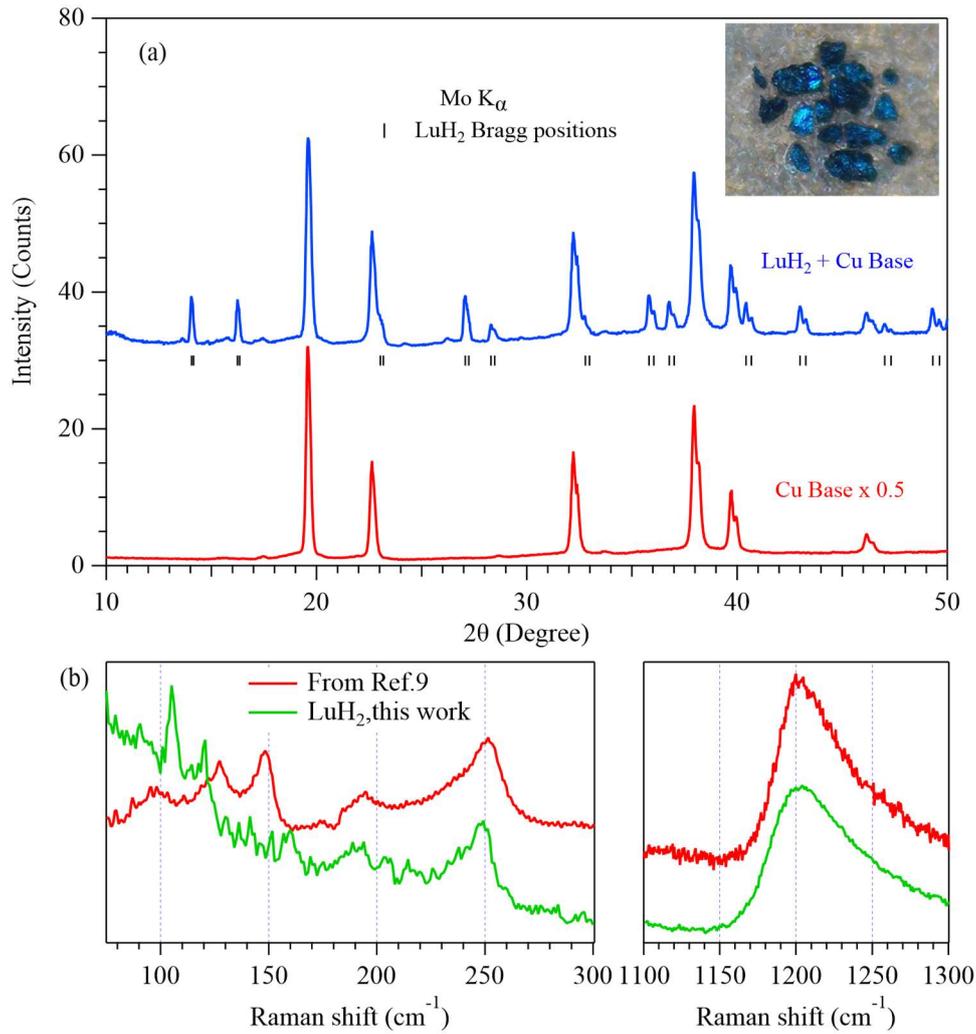

**Figure 2**. (a) The XRD patterns of the LuH$_2$ grains together with the Cu base (top) and the Cu base only (bottom). Inset shows the photograph of the LuH$_2$ grains. (b) The Raman spectra of the LuH$_2$ grain collected at ambient pressure together with that of N-doped Lu hydride from Ref. 9 (Extended data Fig. 1 in Ref. 9) for comparison.

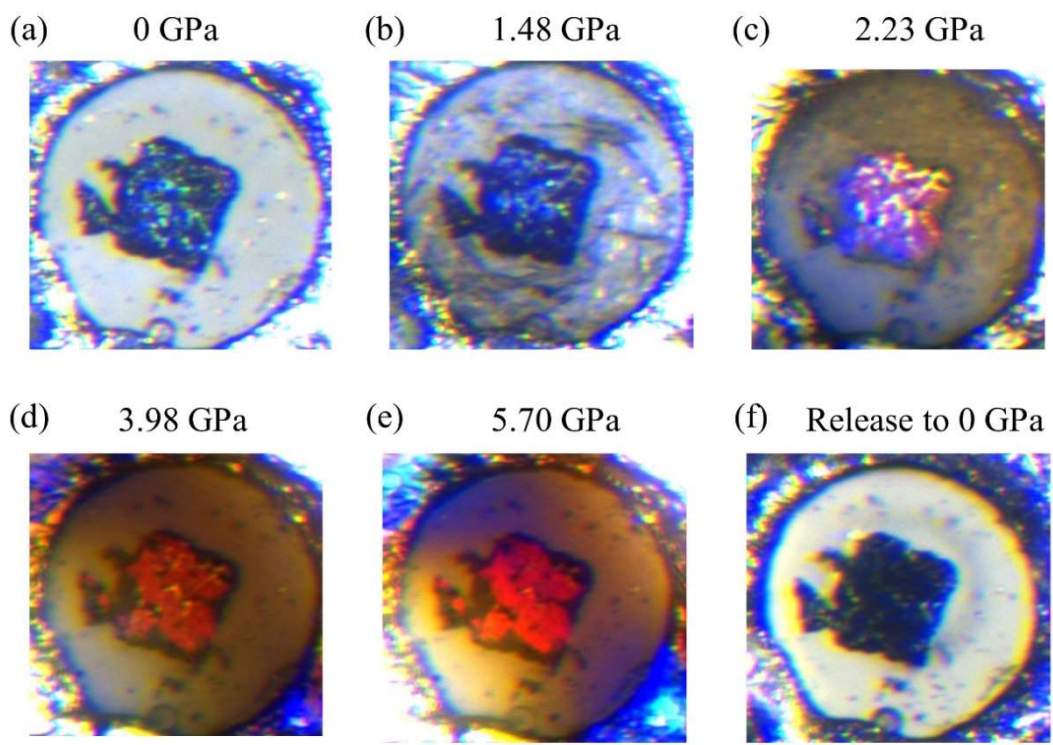

**Figure 3.** The microphotographs of the LuH$_2$ grain with a size of approximately 40 μm × 40 μm in the DAC chamber at different pressures: (a) 0 GPa, (b) 1.48 GPa, (c) 2.23 GPa, (d) 3.98 GPa, (e) 5.70 GPa, and (f) 0 GPa after decompression.

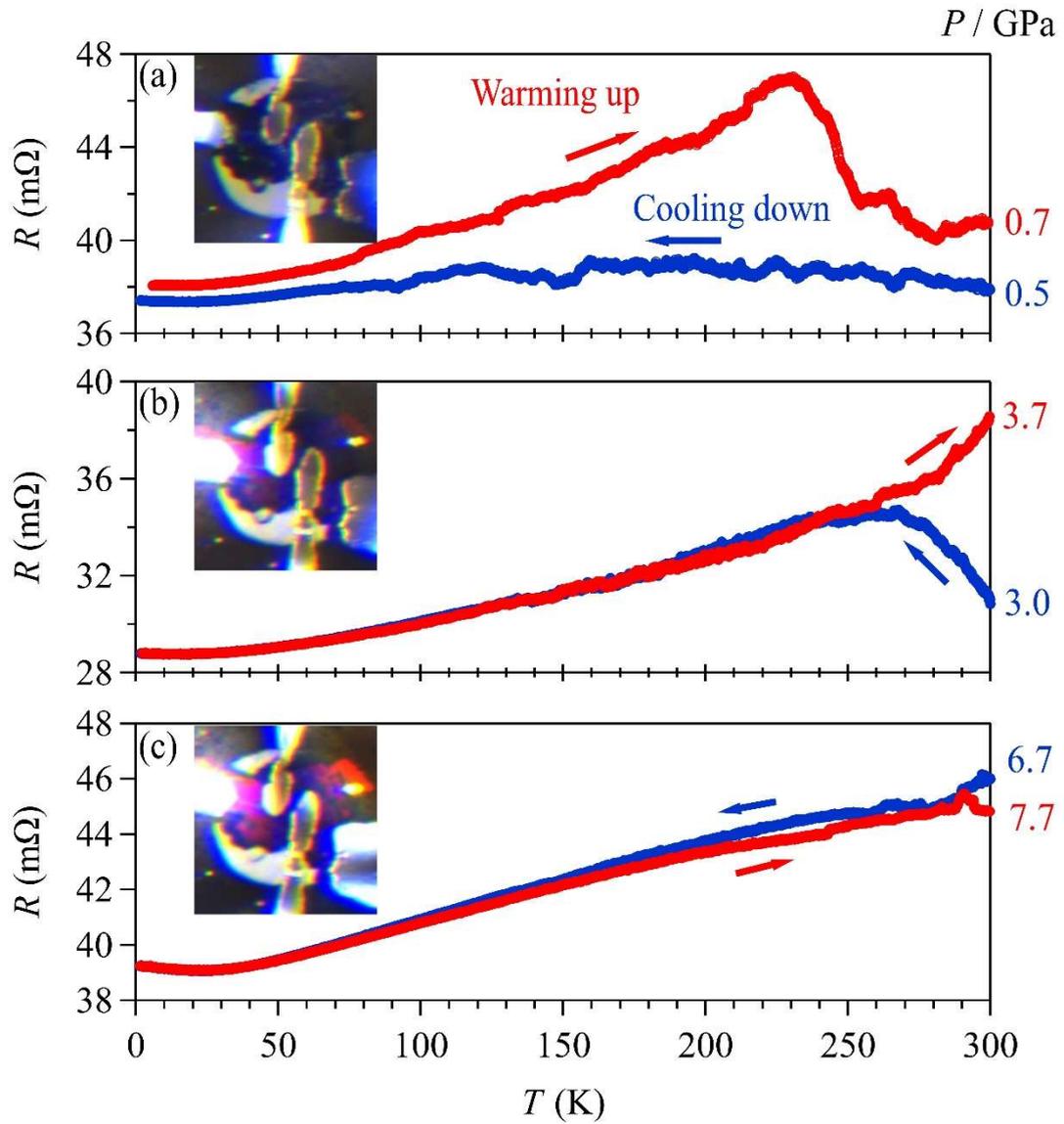

**Figure 4**. Temperature dependence of resistance $R(T)$ of the LuH$_2$ sample at three pressures: (a) 0.5(0.7) GPa, (b) 3.0(3.7) GPa, and (c) 6.7(7.7) GPa. The inset photographs clearly illustrate the color changes of the sample upon compression.

8